\documentclass[12pt]{article}
\usepackage{jheppub}
\usepackage{amsmath,amssymb}
\usepackage{slashed,verbatim,graphicx}
\usepackage{ytableau}

\usepackage{color}

\newcommand\ket[1]{| #1\rangle}

\newcommand\BZ{\mathbb{Z}}

\def\Tr{\textrm{Tr}}

\begin{document}

\title{Negative specific heat from non-planar interactions and small black holes in AdS/CFT.}

\author{David Berenstein}\emailAdd{dberens@physics.ucsb.edu }

\affiliation{ Department of Physics, University of California, Broida Hall, Bldg 572, Santa Barbara, CA 93106}

\abstract{The gravity side of the gauge/gravity duality predicts the existence of small black holes with negative specific heat. 
A free theory of strings has a Hagedorn behavior, but it does not lead to negative specific heat. To understand such states one needs to consider a theory of interacting strings.
In the dual gauge theory, the string interactions are related to non-planar diagrams.
In this paper the simplest gauge matrix model of two free matrices, that has Hagedorn behavior is analyzed in detail. A simple double trace deformation of the Hamiltonian, proportional to square of the free Hamiltonian square with a negative sign that mimics a gravitational attraction is enough 
to produce states with negative specific heat perturbatively and one can still compute the equation of state relating the entropy and the energy. 
A more general argument based on non-planar interactions that are random and that grow faster in strength than the energy suggests that states with negative specific heat appear generically.}

\maketitle

\section{Introduction}

The black hole phase in AdS spacetimes has been argued to be dual to the deconfined phase for a Yang Mills plasma under the gauge/gravity duality \cite{Witten:1998zw}. 
The transition from the deconfined phase to the confined phase is a first order phase transition, which in the gravity setup is reflected in the Hawking-Page phase transition \cite{Hawking:1982dh}. 

The order parameter for the existence of the black hole phase is the change of topology in the Euclidean path integral with fixed  asymptotic boundary conditions:
a circle times a sphere. There are two ways to fill the boundary geometry with different topologies: one where the sphere is filled which corresponds to the thermal AdS geometry (with an energy that is of order one) and one where the circle is filled leading 
to the Euclidean black hole which is like a cigar geometry times a sphere, with an energy of order $1/G \simeq N^2$. Upon analytic continuation to Lorentzian space, the Euclidean black hole goes over to the geometry of the Lorentzian black 
hole. In this geometry one can argue that it is the presence of the horizon that distinguishes the topology. In principle, if one uses the eternal black hole, one has a different topology because one is connecting the system to another copy 
that lives on the other side of the Einstein-Rosen bridge, giving a setup that corresponds to the thermofield doubled state representation of the ensemble \cite{Maldacena:2001kr}. This has been argued to produce the geometry of spacetime via entanglement
\cite{VanRaamsdonk:2010pw}

The existence of a small black hole phase in AdS should still correspond to a deconfined phase of the plasma, because the analytic continuation to the Euclidean theory has the same spacetime topology as a large black hole. 
However,  it is not the dominant configuration in the canonical ensemble.  The small black hole phase in this type of setup has negative specific heat \cite{Hawking:1982dh}. Because these black holes have smaller mass and entropy than the large black holes, it is still possible that this setup corresponds to dual states that are 
dominant (generic)  in the microcanonical ensemble. It is this phenomenon that we want to account for in the gauge theory dual. That is, we want to specify a system where we can count states in the microcanonical ensemble and where we can also account for the negative specific heat.

A rather natural set of questions to ask is what type of physics, at large $N$,  is required in the boundary theory to have such a negative specific heat? Is there a simple model that can accommodate it? More importantly, does that physics require some tuning, or is it generic?
 Such a model might be good enough to address some issues in the relation between black holes and quantum 
information without requiring the precise details of the dual geometry, but a much coarser description. Indeed, one expects in general that many features that appear in gravity setups are much more generic and do not depend on having a weakly coupled gravity dual, but that the weakly coupled gravitational dual makes these effects very apparent.

 Moreover, if the answer is that the physics responsible for such behavior is generic, it would give some hints that some of the features we associate to gravitational physics arise from a universal mechanism in the 
dual theory which is insensitive to the precise details of the dynamics of the boundary theory. The gravity setup would provide a geometrization of such a mechanism.

Rather than talk abstractly about these issues, it is usually best to start from a concrete model. The author recently showed that a simple model of two gauged free $N\times N$ matrices, which has a Hagedorn behavior, can act at finite energy as both a gas of strings and as a gas of excited D-branes giving a simple model that 
indicates the smooth string gas to black hole  transition \cite{Berenstein:2018lrm}, as expected from the works of Susskind, Horowitz and Polchinski \cite{Susskind:1993ws, Horowitz:1996nw}. This occurs at energies that are large with respect to the gap, but small with respect to  $N^2$.
The main idea of the paper is to use this model as a concrete example where one can perform computations. At least when construed as a free model, the counting of states will look like a free gas of strings and the goal is to show that adding certain interactions leads to a computable model where the desired effect, namely negarive specific heat,  
can be calculated in detail.

What is immediately understood from inspecting the problem is that a gas of free strings with Hagedorn spectrum is not enough to produce a negative specific heat in the microcanonical ensemble. If the density of states scales as $\rho(E) \simeq \exp(\beta_H E) E^\alpha$ (in models $\alpha\geq 0$)
\begin{equation}
S= \beta_H E + \alpha \log (E)
\end{equation}
and the inverse temperature is given by
\begin{equation}
\beta= \partial_E S= \beta_H \frac \alpha E .
\end{equation}
This approaches $\beta_H$ from above at high energies. The limit $\beta_H$ is the inverse of the Hagedorn temperature and it indicates  an ultimate temperature of the system.
To achieve a negative specific heat it is necessary that 
\begin{equation}
\partial_E T < 0 
\end{equation}
which is equivalent to $\partial_E^2 S>0$. That means that the entropy must be concave upwards in a computation. The Hagedorn behavior already gives us $\partial_E^2 S\simeq 0$ at large energy, so in principle we need just a small push to make it 
concave upwards. 
The Hagedorn behavior is almost a first order transition. When we have negative specific heat it turn essentially into a first order transition. In many thermodynamics settings the microcanonical and canonical ensembles agree, but not when there is negative specific heat. 
If the model is at infinite volume, one usually has to take care of phase separation to understand the microcanonical ensemble. There can be finite size effects that make the analysis complicated at small volumes.
It is therefore necessary to check in examples to what extent the physics in the canonical and microcanonical ensemble agree.

Consider now such a gas of strings, but let us add a small gravitational attraction. The entropy $S(E_{rest})$ in terms of the rest energy should be the same, but the true energy is redshifted to $E= \exp(-\phi(E_{rest})) E_{rest}$ with some small redshift factor, so that $E_{rest} = E \exp(+\phi(E))$. Substituting we find that
\begin{equation} 
S= \beta_H E_{rest} \simeq   \beta_H  E \exp(+\phi(E))
\end{equation}
and $\partial^2_E S \sim \beta_H  \exp(+\phi(E)) \partial_E\phi(E) >0 $ at small $E$ which indicates that the redshift should increase with the mass of the gas of strings. This is quite natural and suggests that the negative specific heat is rather easy to attain. The gravitational interactions between the strings are due to non-planar
graphs in the dual large $N$ field theory, so  we expect that the main driver of the negative specific heat is a contribution from non-planar diagrams. There are other graphs where in principle, other strings are exchanged, so they could either enhance or conspire against this effect. 
The purpose of this article is to show that such non-planar effects will generically give a negative specific heat in the Hagedorn region and that this does not require tuning the non-planar interactions. Instead,  essentially non-planar interactions that have the statistics of a random matrix distribution do that.

However, the first goal is to deform the model of two free matrices slightly with a correction that is soluble and essentially non-planar. The simplest such correction is a double trace deformation that is scaled appropriately for the large $N$ limit. To make it soluble, we use the free Hamiltonian itself, with a negative sign, to mimic an effect that is similar to gravitational attraction. This will show how gravity can be heuristically {\em solely} responsible for the effect we want. I will then show that a more sophisticated version of the non-planar contributions seems to {\em generically} point in the direction of producing a negative specific heat. The idea is that non-planar diagrams should give some form of a corrected effective Hamiltonian (at roughly fixed energy) that should be similar to random matrices. This is with the understanding that generic chaotic systems have the same statistics between their levels as random matrices. That is, one can assume that the matrix elements of a chaotic hamiltonian are like a random matrix  in whichever preferred basis we have and a computation shows that the main effect is to make the second derivative of the entropy with respect to the energy positive.

The paper is organized as follows. In section \ref{sec:can} the canonical ensemble of the free matrix model with two matrices, in the presence of a chemical potential is analyzed and the details of the limit temperature curve as a function of the chemical potential are analyzed.
In section \ref{sec:micro},  bounds on the micro-canonical counting of states at fixed charge and energy are obtained. This serves as a specific example to compare to the canonical setup later on. Section \ref{sec:thermo} derives the thermodynamic quantities and relations among them from the previous results. It is shown that this matches the canonical ensemble critical curve. Section \ref{sec:beyhag} deals with the interpretation of the system as a black hole, and with deviations from the Hagedorn behavior that are introduced by non-planarities in two setups. The first one is a double trace deformation, while the second one is a more generic model of `random' non-planarities. Section \ref{sec:beytwo} deals with gauged matrix models with more matrices.  Finally, in \ref{ref:con}  I conclude.

\section{The canonical ensemble}\label{sec:can}

Consider a gauged matrix quantum mechanics for two free matrix degrees of freedom, $X,Y$, each with the same mass $M=1$. The canonical partition function with chemical potential $\mu$ at infinite $N$ can be expressed as follows \cite{Aharony:2003sx} (see also \cite{Sundborg:1999ue})
\begin{equation}
Z[\beta, \chi] = \prod_{k=1}^\infty \frac1{1-e^{-\beta k} e^{\chi k /2} -e^{-\beta k} e^{-\chi k /2}  } =\prod_{k=1}^\infty \frac{1}{1-x^k-y^k}
\end{equation}
where $\chi= \beta \mu$, and we assume that $X$ has charges $1/2$, while $Y$ has charge $-1/2$. It is natural to assign these charges because the system has an $SU(2)$ symmetry. It is more convenient to use the  
fugacity $\chi$ than the chemical potential itself \footnote{More precisely, the fugacity should be $\exp(\chi)$, which counts the charge. We can also say that this is a normalized chemical potential.}. Also, we can use the $x,y$  variables defined by $x, y= \exp(-\beta) \exp(\pm\chi/2)$ which count the number of $X,Y$ respectively and these variables are both positive. 

The fact that the partition function itself diverges at $x+y=1$ indicates that there is a maximum temperature: the partition function does not make sense beyond this temperature. The divergence is localized in the first term of the product. When $\chi=0$ this is the Hagedorn behavior associated to confinement/deconfinement when we go to finite $N$. The transition curve is given by
\begin{equation}
\exp(\beta) = \exp(\chi /2) +\exp(-\chi/2)\geq 2
\end{equation}
So the minimum value of $\beta$ is $\log(2)$ and this implies a maximum temperature $T_{max}= \beta^{-1}= 1/\log(2)$. 
The quantity $x+y$ can be thought of as a one site partition function for a single choice of the letter $X,Y$. At criticality, $x/(x+y)=x$ and $y/(x+y)=y$, so $x,y$ can be interpreted as probabilities themselves. 
The main properties of the partition function are little changed if we use fermions instead of bosons, or if we mix one fermion and one boson, as the Hagedorn temperature will appear in the same place. We will not pursue these directions further, although if one wants to, one can also consider a supersymmetric system with these properties \cite{Raha:2017jgv,Curtright:2017pfq} (see also \cite{Beccaria:2017iqb}).

Given $\chi$, the probability of choosing an $X$ over a $Y$ in a given configuration is
\begin{equation}
p_x= \frac{\exp(\chi/2)}{\exp(\chi /2) +\exp(-\chi/2)}
\end{equation}
whereas 
\begin{equation}
p_y= 1-p_x= \frac{\exp(-\chi/2)}{\exp(\chi /2) +\exp(-\chi/2)}
\end{equation}
At energy $E$, the expected number of $X$  in the state is $N_x=p_x E$, and the expected number of $Y$ is $p_y E$. When $\chi=0$ is the values at which $X$ versus $Y$ are equally probable.
The charge of the state at energy $E$ from the canonical ensemble is therefore
\begin{equation}
J= \frac 12 (N_x-N_y) =  \frac E2 \tanh(\chi/2) 
\end{equation}
One of the goals of this paper is to show that the canonical and microcanonical descriptions of the system agree, so it is necessary to have this information on hand to compare them.

\begin{figure}[ht]
\begin{center}
\includegraphics[width=8cm]{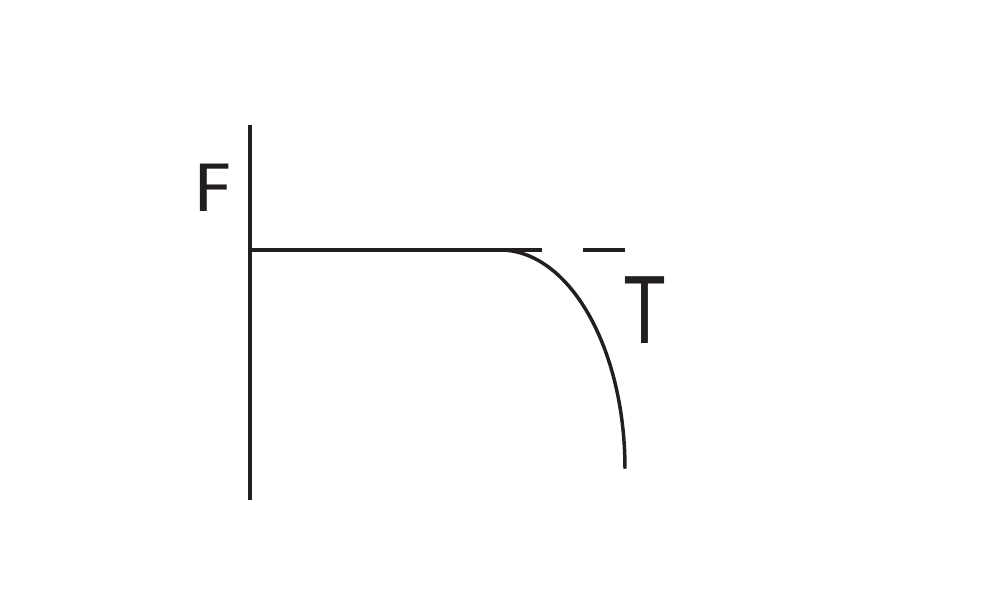}\caption{Free energy versus temperature in the canonical ensemble}\label{fig:FvsT}
\end{center}
\end{figure}

When we move to finite $N$, but $N$ large,  the physics is different. The main reason is that the Hagedorn temperature is not the ultimate temperature of the system any longer. if one uses a path integral formalism with a Polyakov loop in the Euclidean setup, one finds that the free energy, as a functions of temperature,  vanishes up to the Hagedorn temperature. After that temperature is reached the system acquires a positive specific heat again and the entropy grows asymptotically like $N^2 \log T$, whereas the energy grows linearly in $T$. The free energy is negative. At this critical temperature, the theory shows non-trivial critical exponents when approaching the critical temperature from the top. This can be captured simply by a graph of the free energy versus temperature in figure \ref{fig:FvsT}.

 The temperature where the free energy starts going negative is the Hagedorn temperature. At that value there are non-trivial critical exponents, which are associated to the Gross-Witten phase transition \cite{Gross:1980he} being on top of the Hagedorn proliferation of states \cite{Aharony:2003sx}. It is natural to associate this
 additional transition to the finite $N$ nature of the physics. This transition should broadly indicate that there will be relations in the counting of states so that the entropy becomes reduced from the Hagedorn behavior. This  in turns leads to being able to increase the temperature further, beyond the Hagedorn temperature.
 Notice that the Hagedorn density of states $\rho(E) \simeq E^\alpha \exp(\beta_H E)$ enters in the partition function as follows
 \begin{equation}
 Z[T] \simeq \int dE \rho(E) \exp(-\beta E)
 \end{equation}
For convergence we need $\beta < \beta_H$. To have convergence at $\beta <\beta_H$, we need that at asymptotically large $E$ we have that  $\rho(E) < \exp(\beta E)$ for all $0<\beta<\beta_H$. This indicates that the density of states and hence the entropy, is very reduced.
These relations should dominate in the large temperature regime, so that a naive gas of quasi free strings is the wrong picture to describe the states. In the high temperature phase, the $U(N)$ should be deconfined instead and in the analogy to black holes, this is the large black hole phase that has positive specific heat.
In this paper we are interested mostly in the {\em small black hole} phase, which occurs in a different part of the phase diagram.

\section{Microcanonical Counting state inequalities}\label{sec:micro}

We now want to analyze the system (at large $N$) in a microcanonical ensemble at fixed energy and fixed charge. Our problem is to count states in the field theory carefully.
The first simplest idea is to count the single trace states (we will call these single string states). A single trace can be thought of as a state
\begin{equation}
\ket \psi = \Tr(X Y X\dots)\ket 0
\end{equation}
Because the letters inside the trace are ordered, we can think of this as counting words made of $X,Y$. We want to count them at fixed numbers of $X$ and $Y$. 
In this counting, we have a total of $E$ sites on the trace, and out of these $M$ are $X$, and $E-M$ are $Y$. There are 
\begin{equation}
Q_{word}= {E \choose M}\label{eq:word}
\end{equation}
such words. 
However, the cyclic property of the trace tells us that there is an identification on these words by the group action of the cycle. This group action is $\BZ_E$ and is of order $E$
 If the group action is free on these words, there are only $Q_{1-string}= Q_{word}/E$ such words. If the action is not free, when we divide we are undercounting because for some words there are fewer images and we should divide by the subgroup of $\BZ_E$ that acts freely on the word. 
 We get this way our first inequality
 \begin{equation}
Q_{1-string}\geq \frac 1E {E \choose M}
\end{equation}
Similarly, if all strings are fixed points, we should not divide at all and we get
 \begin{equation}
Q_{1-string} \leq {E \choose M}
\end{equation}

Now, we want to count multitrace states similarly.

Notice first, that in a multi-trace state, we can order the traces so that the length of the words in each trace is in ascending (or descending) order.
With this ordering, we can put the words inside each trace in consecutive order and obtain one long word, finding the same number as in \eqref{eq:word}.

Just like in the case of a single string, there will now be a cyclic action on each trace. But more than this, there is also a permutation group action that permutes traces with the same number of letters (strings of the same length).
A multitrace therefore consists of a partition  $\pi$ of $E= \sum n_k$. The group action $\Gamma_\pi$ has 
\begin{equation}
|\Gamma_\pi| = \prod_k n_k! k^{n_k}
\end{equation}
elements. If $\Gamma_\pi$ acts freely, we find that for mulitraces associated to the partition $\pi$ we get
$Q_{\pi-string}= Q_{word}/|\Gamma_\pi|$ states. Again, some of these are undercounted because the group $\Gamma_\pi$ does not generally act freely. For example if all $X$ are in one trace and all $Y$ in another, then the group leaves the word fixed. We find then that there is an inequality
\begin{equation}
Q_{\pi-string} \geq Q_{word}/|\Gamma_\pi|
\end{equation} 
and similarly to the one string case $Q_{\pi-string}\leq Q_{word}$.
Summing over the partitions $\pi$ we get that 
\begin{eqnarray}
Q_{multi-string}= \sum_\pi Q_{\pi-string} &\geq& \sum_\pi Q_{word}/|\Gamma_\pi|\\
&\leq & P(E) Q_{word}
\end{eqnarray} 
where $P(E)$ is the number of partitions of $E$.

We get this way that the total number of multitraces is bounded below and above by
\begin{equation}
{E \choose M} \sum_{\{n_k\}\vdash N} \frac 1{\prod_k n_k! k^{n_k}} = {E \choose M} \leq Q_{multi} \leq P(E) {E \choose M} 
\end{equation}
This is already enough to do thermodynamics in the large $M,E$ limit whit $M/E$ fixed.
 
Notice that above we have assumed that the strings are free: we have not imposed any relations between traces as would be expected at finite $N$. Our only concern has been to count multitraces up to cyclicity of each trace and permutations of identical strings.

 \section{Thermodynamics in the microcanonical ensemble}\label{sec:thermo}
 
 One naturally expects that the following is true in the large $E,M$ limit, by using Stirling's approximation
 \begin{eqnarray}
{E \choose M} &\simeq& \exp( E\log E-M\log(M)-(E-M)\log((E-M))) \\
&=& \exp( -M\log(M/E)-(E-M)\log((E-M)/E))\\
&=&   \exp( E s)
 \end{eqnarray}
 where we define the one site entropy
  \begin{equation}
  s= -M/E\log(M/E)-(E-M)/E\log((E-M)/E)) = -p_x\log p_x-p_y \log p_y
  \end{equation}
   we also have the asymptotic behavior $P(E) \simeq \exp(O(\sqrt E))$

 From here we get that 
 \begin{equation}
\exp( E s) \leq  Q_{multi} \leq  P(E) \exp( E s)
 \end{equation}
where $p_x$ and $p_y$ are the probabilities that an individual letter in a word is $X$ or $Y$.

Taking logarithms 
\begin{equation}
E s \leq \log Q_{multi} \leq E s + O(\sqrt E)
\end{equation}
and the term of order $\sqrt E$ is subleading with respect to $E$. 

Similarly, 
\begin{equation}
E s -log(E) \leq \log Q_{1-string} \leq E s 
\end{equation}
 so that the 1-string ensemble is just as good a the multi-string ensemble to get the leading order entropy computation. 
 
The associated large $E$ entropy is
 \begin{eqnarray}
 S(E,M) &=&  -M\log(M/E)-(E-M)\log((E-M)/E) \\
 = S(E,J) & =& - \left( \frac E 2 + J\right) \log\left(\frac 1 2 + \frac J E\right)- \left( \frac E 2 - J\right) \log\left(\frac 1 2 - \frac J E\right) \label{eq:entje}
 \end{eqnarray}
where we have introduced the charge 
\begin{equation}
J= (N_X-N_Y)/2 = M -E/2
\end{equation}
 
 Using the first law of thermodynamics
 \begin{equation}
T dS = dE - \mu dJ 
\end{equation}
 we get that 
 \begin{equation}
 \beta = \frac{\partial S}{\partial E}|_J
 \end{equation}
 and that
 \begin{equation}
 \beta \mu =\chi= -\frac{\partial S}{\partial J}|_E
 \end{equation}
 
 From here we get that the inverse temperature is
 \begin{equation}
 \beta = \frac{\partial S}{\partial E}= 1/2 (-\log(1/2 - J/E) - \log(1/2 + J/E))\label{eq:temp}
 \end{equation}
 whereas the fugacity is 
 \begin{equation}
 \chi= \frac{\partial S}{\partial J}  = \log(1/2 + J/E)-\log(1/2 - J/E) 
 \end{equation}
 It  is easy to verify explicitly that
 \begin{equation}
 \exp(\beta) = \exp(\chi/2)+\exp(-\chi/2)
 \end{equation}
 which reproduces the critical temperature curve of the canonical ensemble. We also find that
 \begin{equation}
 \frac{N_x}{N_Y}= \frac{1/2 + J/E} {1/2 - J/E} = \exp(\chi)= \frac{p_x}{p_y}
 \end{equation}
 which reproduces the results from the canonical ensemble at criticality relating the fugacity to the probabilities of $X$ versus $Y$ sites.

\subsection{Beyond Hagedorn}

\begin{figure}[ht]
\begin{center}
\includegraphics[width=8cm]{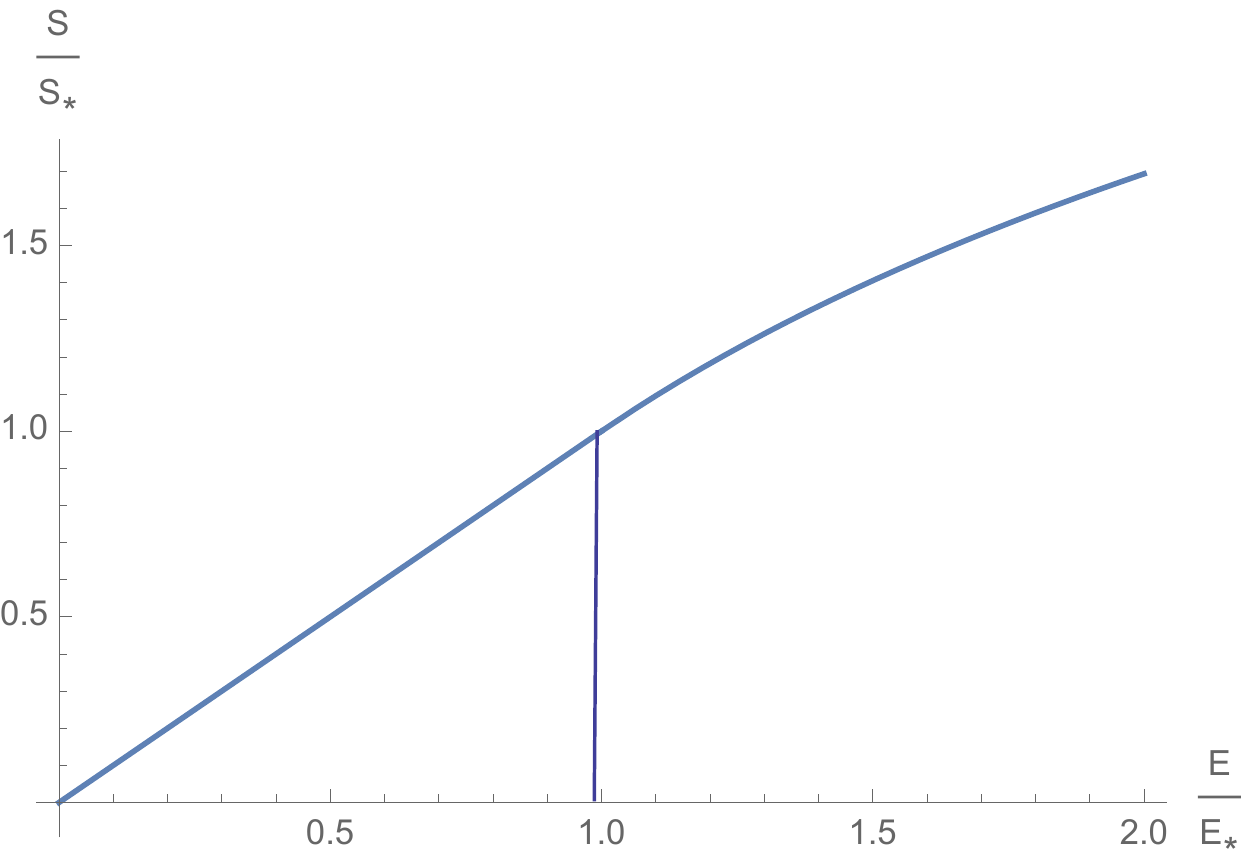}\caption{Entropy versus Energy. The plot has a continuous phase transition at $E_*$, where the slope is continuous but there are non trivial critical exponents at $E_*$ on the right hand side. }\label{fig:SvsE}
\end{center}
\end{figure}

Consider now the exit of the Hagedorn behavior. 
This will occur at a large energy $E_*$ which is of order of $N^2$. Because for Hagedorn behavior we have that $S= \beta_H E$, we also have that the corresponding entropy 
$S_*$ scales like $N^2$. When we go above the energy $E_*$ we expect the entropy to scale slower than $E_*$ so  that we can start increasing the temperature, using the thermodynamic 
relation $\beta = \partial_E S$, but we expect $\beta$ to be continuous at this point. The main reason for $S$ to start behaving differently has to do with finite $N$. It must be the case that at this point 
some relations between traces are becoming apparent. Such relations would be akin to a stringy exclusion principle \cite{Maldacena:1998bw}.  For matrix models of a single matrix, the identities between traces start at length $N$ and there 
is an interpretation of the stringy exclusion principle in terms of extended objects  represented by D-branes becoming relevant \cite{McGreevy:2000cw}. 
Obviously it is interesting to understand the physics of $E_*$ and $S_*$ near the transition. We expect the phase diagram to look as in figure \ref{fig:SvsE} with non-trivial critical exponents on the right hand side of the transition.
This is what we would expect if the physics at the singularity is the same physics of the figure \ref{fig:FvsT} in the thermodynamics variables $S,E$ rather than $F,T$.
To understand the relations between states we would need to go to a different basis. Such a basis is provided by the Young diagram description of the states. It in this new basis that a description in terms of D-branes becomes more apparent \cite{Berenstein:2018lrm}. It is also in this description that giant
 gravitons can be matched to specific states \cite{Balasubramanian:2001nh}, and the stringy exclusion principle is then the statement that representations of $SU(N)$ can have at most $N$ rows. For half BPS states, this map between states and giant gravitons is well understood \cite{Corley:2001zk}.
 A proper understanding of this transition in the multimatrix case requires us to know what the ``typical state" looks like below $E_*$ and how that gets modified at $E_*$ (this is currently under investigation \cite{BR}). The natural guess is that the corresponding typical Young diagrams are growing so that they start touching the maximum height allowed. 
 However, these effects turn the specific heat positive again. Any hope of having a negative specific heat must lie elsewhere and must include additional interactions.

 \section{Black hole interpretation}\label{sec:beyhag}

 Now let us turn to the interpretation of the system as a small black hole, expanding on the ideas in \cite{Berenstein:2018lrm}. This is not to say that the system is a black hole, but rather, the purpose is to study which characteristics of black holes 
 are readily apparent in this example and to compare to such setups, that is we want to know how black-hole like the system is. Since we have been working in the microcanonical ensemble at fixed energy and 
 charge we can analyze the black hole temperature as a function of the charge at fixed energy.
 
\begin{figure}[ht]
\begin{center}
\includegraphics[width=8cm]{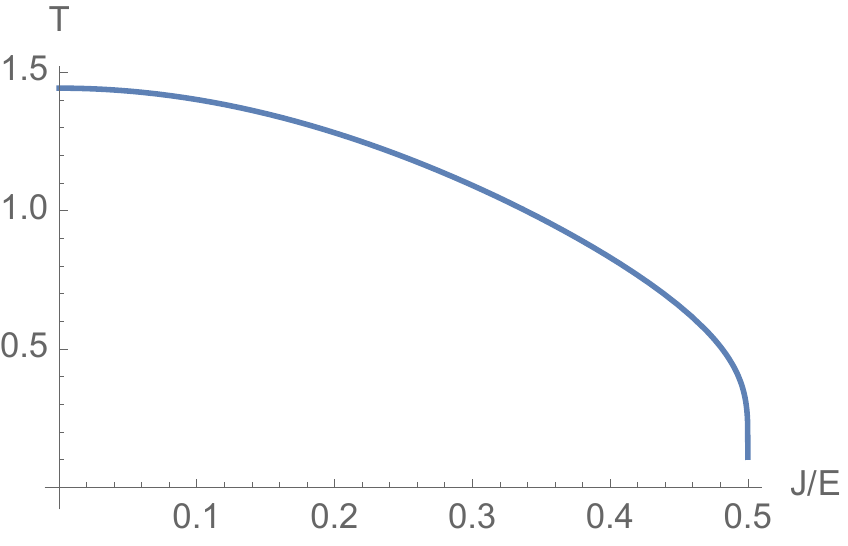}\caption{Temperature versus charge. The plot is symmetric on $J\to -J$, so negative $J$ is not shown.}\label{fig:TvsJ}
\end{center}
\end{figure}

 Notice that in our calculations in equation \eqref{eq:temp} we get that the temperature is given by
 \begin{equation}
 T= \frac{2 }{(-\log(1/2 - J/E) - \log(1/2 + J/E))}
 \end{equation}
The temperature only depends on $J/E$. The temperature versus charge curve is show in figure \ref{fig:TvsJ}.
 Notice that we have $J/E\leq 1/2$ from how the charges of the system are arranged: the maximal charge is if we have $N_x=E$, and this state has charge 
$J=1/2E$. In the temperature computation on nearing the bound we have that the term $\log(1/2-J/E)$ diverges and the temperature tends to zero. 

In this limit, the entropy is subleading in the energy. In particular, we find that at fixed $E$ and on saturating the charge, there is still a non-zero entropy. This is 
because the system reduces to a one matrix model. In a one matrix model the number of states at fixed $E$ is equal to partitions of $E$. The entropy only grows as $\sqrt E$
and not linearly in $E$ (we need to do the calculation at fixed extremality parameter $j= J/E=1/2$).

This is as would be expected for an extremal black hole: extremal black holes are characterized by a degenerate horizon so that the temperature of those black holes vanishes.
If we naively continue past this point we end up in a situation with a complex temperature and complex entropy, which is unphysical. Such states are not part of the Hilbert space of states. One 
can assume that the complex temperature/entropy is a signal similar to the existence of closed timelike curves in certain solutions of supergravity: they indicate a broad violation of unitarity \cite{Caldarelli:2004mz,Milanesi:2005tp}.

\subsection{The problem of negative specific heat}
 
Unlike small black holes in AdS, the specific heat is singular in that $C\simeq (\partial T/\partial E)^{-1}$, because the temperature is constant,  but the specific heat never goes negative. This is expected of a free gas of strings. Indeed,
so long as a single long string dominates the ensemble, both the energy and the entropy are proportional to the length of the string that is measured on the number of letters in the gauge invariant word, or similar planar description. 
The theory on the string is local, so the energy density and the entropy density should be well defined. To go beyond this regime, one needs a source of nonlocality that sources interaction between different parts of the string, or between 
different strings. These should be controlled by non-planar diagrams.

We expect in general that if we perturb the system away from free, the reason to turn the specific heat negative in the microcanonical ensemble should be related exclusively to non-planar diagrams.
This possibility will turn the Hagedorn transition from a weak first order transition to a first order first transition in the canonical ensemble. 

To understand this,  consider the canonical ensemble at fixed temperature $T=\beta^{-1}$. Then we have that 
\begin{equation}
Z[\beta]= \int dE \exp(-\beta E+S(E))
\end{equation} 
where $S(E)$ is the entropy in the microcanonical ensemble. If we view this integral as a contribution from saddles, the condition to get a saddle is that 
\begin{equation}
\partial_E(-\beta E+S(E))=0 = -\beta +\partial_ES
\end{equation}
This last condition is the {\em definition} of the temperature in the microcanonical ensemble. To have a good saddle, we should also have that 
\begin{equation}
\partial_E^2 S <0 
\end{equation}
This is equivalent to 
\begin{equation}
\frac{\partial \beta}{\partial E} <0 = -\frac 1 {T^2} \frac{d T}{d E} = -\beta^2 C^{-1}
\end{equation}
where $C$ is the specific heat. That is, the condition for a proper saddle is that the specific heat is positive.
When these conditions occur, if the saddle is very pronounced, the microcanonical and the canonical ensemble give the same answer. 
This is usually accomplished by taking the thermodynamic limit. In matrix models, this happens when $N\to \infty$. 

When this does not occur, the saddle point in question is either at a minimum rather than a maximum, 
 or, the system is degenerate if the heat capacity is divergent. This is the case with 
the Hagedorn gas, where in the microcanonical ensemble we have that $S$ is linear in $E$ and therefore the second derivative vanishes.

If the heat capacity goes negative, we expect that there are other maxima that will dominate in the canonical ensemble. One of these will have the smallest free energy 
\begin{equation}
F= E-T S
\end{equation}
and this will be the dominant configuration at the given temperature. At this saddle the specific heat is positive.
The switching from one dominant local maximum to another one will be the first order phase transition as understood in the canonical 
ensemble. 
The region of negative specific heat therefore does not lead to a configuration that is important in the canonical ensemble.

However, in the microcanonical ensemble, this region can dominate the dynamics. To get negative specific heat we need the equivalent of the figure \ref{fig:SvsE} to be concave 
upwards in some region.  In principle a small perturbation can do that already in the region where the entropy is linear in the energy, and that is at energies that are smaller than $E_*$. 
We can not do that with a small perturbation beyond $E_*$, as we would need to move the second derivative $\partial_E^2S$ by a finite amount in that region, rather than by an infinitesimal amount.

Let me give an example of a very simple solvable phenomenological model perturbing away from free that can capture this possibility.
Consider the Hamiltonian (acting on words) given by
\begin{equation}
\hat H = \Tr( X\partial_X) + \Tr( Y\partial_Y) = \hat T.
\end{equation}
This way of writing the hamiltonian is useful when one treats $X$ as a (matrix) raising operator and $\partial_X$ as the lowering operator. The letter $T$ should remind us of the stress tensor.

Now, consider a modified  model given by
\begin{equation}
\hat H' = \hat T - \frac{\beta}{N^2} \hat T^2\label{eq:modify}
\end{equation}
where $\beta$ is a small parameter of order one.  The operator $\hat T$ roughly counts mass and an operator of the form $ \hat T^2$ with negative coefficient
gives an effect that is similar to the attraction of gravity: it is proportional to the mass squared. This is a double trace deformation and acts like non-planar diagrams, as the two actions of
$\hat T$ do not have to sit near each other in a word in order to contribute. The factor of $1/N^2$ is akin to the gravitational constant in a putative dual geometry.

A better deformation is given by
\begin{equation}
\hat H' = \hat T - \frac{b}{N^2} \frac{\hat T^2 }{1+(\gamma/N^2) \hat T}
\end{equation}
which has the advantage of not turning negative for very large $\hat T$ so long as $\gamma$ is large enough, but at small enough $\hat T$ it makes little difference.
This is a finely tuned setup in that it still allows us to compute the degeneracy of states at fixed $\hat T$: it is the same as before.

Also, remember that double trace deformations should not affect the theory of gravity except for the boundary conditions at infinity \cite{Witten:2001ua,Berkooz:2002ug}, so we do not expect to have introduced 
a severe non-locality in the bulk. We should be able to interpret this as gravitational physics as well.

In the new microcanonical ensemble, fixing $\hat H'$ is equivalent to fixing $\hat T$. We have already computed the entropy at fixed $\hat T$, finding that $S\propto \hat T$. What that means is that
we can write a new relation between the energy and the entropy at entropies that are of order a small number times $N^2$ giving rise to 
\begin{equation}
E \simeq \alpha S -\frac{\alpha^2b}{N^2} S^2
\end{equation}
inverting the relation we get that 
\begin{equation}
S= \alpha^{-1}  E  +\frac{ b}{\alpha N^2} E^2
\end{equation}
In this new setup we have that 
\begin{equation}
T(E) \simeq \frac1{\alpha^{-1} + \frac{2b}{\alpha N^2} E }
\end{equation}
and $T$ decreases as $E$ increases, indicating a negative specific heat at intermediate energies in the microcanonical ensemble.
This is exactly what we wanted.

Another way to do the computation is the following. Consider the entropy formula \eqref{eq:entje}, but with $\hat T$ instead of $E$. 
\begin{equation}
S= - \left( \frac {\hat T} 2 + J\right) \log\left(\frac 1 2 + \frac J {\hat T}\right)- \left( \frac {\hat T} 2 - J\right) \log\left(\frac 1 2 - \frac J {\hat T}\right)
\end{equation}
From equation \eqref{eq:modify}
we get that $E$ is function of $\hat T$, or equivalently, $\hat T$ is a function of $E$. We then get that  the inverse temperature is
\begin{equation}
\beta = \frac{\partial S}{\partial E} = \frac{\partial S}{\partial \hat T} \frac{\partial \hat T}{\partial E} = \beta_{eff} \frac{\partial \hat T}{\partial E}
\end{equation}
where $\beta_{eff}$ is an effective temperature assuming  ``no gravity", while
 the function $\frac{\partial E}{\partial \hat T}<1 $ can act in a similar way to a redshift factor. 
 
 Notice that the extremal black holes that still have $\beta_{eff}\to \infty$.  So they will still have zero temperature and behave as expected in a gravitational dual.

\subsection{Negative specific heat is generic}

Let me now assert that negative specific heat is generic. To make this assertion, I'm going to consider more {\em generic} non-planar corrections. The model I will analyze is based on two assumptions.
First, in the two matrix model I'm going to assume that the number of letters in a trace $L$ is fixed by the interactions (this is not strictly necessary), but we need a counting parameter that is not the energy itself to keep track of the states.
For example, there are deformations of the Hamiltonian that preserve $L$ and $SU(2)$. For example consider
\begin{equation}
\delta H = g_{YM}^2\Tr( [X,Y][\partial_X,\partial_Y])
\end{equation}
This deformation computes the one-loop anomalous dimension in the $SU(2)$ sector of ${\cal N}=4 $SYM and the planar physics is solved by the $XXX$ Bethe ansatz \cite{Minahan:2002ve}.
One can consider other systems of similar type, like those that arise in the works \cite{Harmark:2014mpa, Harmark:2016cjq}  on spin matrix theory. They are defined by a quartic interaction that 
commutes with the number of letters in a trace.

This is just a crutch to make the model more easily tractable, but it should be possible to dispense with this assumption. We can equally assume that $L$ is related linearly to the planar energy of a state. 
 Secondly, I'm going to assume that the
non-planar interactions lift the degeneracy the same way as a random matrix would. This is line with the idea that generic systems are non-integrable and that their quantum dynamics is similar to that based on a random Hamiltonian.
As such, states at level $L$ will get a spread that has a semicircle distribution, which I will parametrize by $x$, and a width that scales with $L$ as $L^{1+\epsilon}$ with a small coefficient $\delta$. 

These assumptions permit one to do a computation and are somewhat reasonable. However, it might be the case that in a more realistic setup the non-planarities have a finer structure than the appearance of randomness. Similarly, we might not have a parameter like $L$
and it is not clear how generic these assumptions really are. It might be that these assumptions are not realistic with respect to

Now, I want to evaluate the density of states at fixed energy $E$. This will look as follows
\begin{equation}
\rho(E) = \int dL \int_{-1}^1 dx \sqrt{1-x^2} \exp(s_0 L)\delta(E-\gamma L +x \delta L^{1+\epsilon})
\end{equation}
where we see that the energy is $\gamma L$ plus a random matrix correction proportional to $L^{1+\epsilon}$. I make no assumption yet on $\epsilon$, but I am assuming that the correction strength is small  (this is where the $1/N$ counting is being assumed).
The sign of $\delta$ can be taken positive: we take $x\to -x$ to compensate without changing the integral.
The factor $\sqrt{1-x^2}$ is the semicircle distribution. What is important is that it has endpoints, not that it is a semi-circle.
Now let us solve the integral by doing the integral over $L$  first, at large $E$, but much smaller than $N^2$, so that we are clearly in the regime where long strings are populated with high probability.
We then get that 
\begin{equation}
E-\gamma L +x \delta  L^{1+\epsilon}=0
\end{equation} 
so that $L\simeq E/\gamma+O(\delta) $. 

 Substituting this value in the second small term, we get that 
\begin{equation}
L = \frac{E}{\gamma} +\frac{ x \delta}{ \gamma} \left(\frac{E}{\gamma}\right)^{1+\epsilon}
\end{equation}
We find this way that roughly
\begin{equation}
\rho(E) = \int_{-1}^1 dx\sqrt{1-x^2}  \exp\left(s_0\left[     \frac{E}{\gamma} +\frac{ x \delta}{ \gamma} \left(\frac{E}{\gamma}\right)^{1+\epsilon}\right]\right)
\end{equation}
where we are ignoring the measure term from the $\delta$ function evaluation. We want to be at large enough $E$ so that the term on the right hand side of the exponential is much bigger than one in absolute value. That is, 
we want to have that 
\begin{equation}
s_0\frac{ \delta}{ \gamma} \left(\frac{E}{\gamma}\right)^{1+\epsilon}>>1
\end{equation}
While still be much smaller than $E$.
 \begin{equation}
\frac{s_0 E}{\gamma }>> s_0\frac{ \delta}{ \gamma} \left(\frac{E}{\gamma}\right)^{1+\epsilon}>>1
\end{equation}
which occurs for a medium size $E$ for small $\delta$. The parameter $\delta$ is the one that scales like $1/N^2$ when $\epsilon=1$.

  The point is that this term grows with $E$.
 Then we get that the integral over $x$ is dominated by
the largest value at $x=1$. We get this way that 
\begin{equation}
\rho(E) \simeq \exp\left(s_0\left[     \frac{E}{\gamma} +\frac{  \delta}{ \gamma} \left(\frac{E}{\gamma}\right)^{1+\epsilon}\right]\right)
\end{equation}
This shows that the entropy is corrected to
\begin{equation}
S(E) \simeq s_0 \frac{E}{\gamma}+ s_0 \frac{  \delta}{ \gamma} \left(\frac{E}{\gamma}\right)^{1+\epsilon}
\end{equation}
with $\delta/\gamma>0$. 
The first term on the right hand side is the usual Hagedorn density of states. The theory will have negative specific heat if $1+\epsilon>1$. That is, if the corrections due to non-planar interactions grow with energy faster than the energy.
This is expected on general grounds: non-planar interactions usually grow like the total possible number of contractions in a collection of words, rather than like consecutive contractions on a single word. The latter ones only scales like the length of the word.

The main mechanism that is at play here is that more states lower their energy from above and by a larger amount than those that increase their energy from below. The gain is mostly because of entropy reasons and this is enough to tilt the transition into a first order 
phase transition. Notice that the effect is sufficiently universal that it is insensitive to a lot of details in the Hamiltonian. In this sense, one can think of this result as a universal `attraction' that could be related to how gravity could emerge from thermodynamics in these models.
Also notice that most of the discussion only depends on a Hagedorn density of states and a notion of nonplanar corrections. In this sense, it is not tied to having only two matrices and it could equally apply to the ${\cal N}= 4$ SYM theory in $3+1$ dimensions.

For more specific models, one can show self-consistency of the negative specific heat with other assumptions about the dynamics, as in the work of \cite{Hanada:2016pwv}. There it is argued that submatrix degrees of freedom are in a deconfined phase and the rest in another and one can get 
qualitative agreement with the small black hole dynamics in AdS.  This is an argument of a very different nature than the one, as it depends on some other details of the dynamics. 
This type of configuration was originally advocated in \cite{Asplund:2008xd}, where a model with correlations between eigenvalues of matrices and position on the sphere was being studied.  above.

\section{Beyond two matrices}\label{sec:beytwo}
 
 It is easy to generalize some of the results above to the case  where we have more than two free matrices, let us say $s$. For the canonical partition function we consider the large $N$ product form \cite{Aharony:2003sx}
 \begin{equation}
 Z[x_\alpha] = \prod_k \frac 1{1-\sum_\alpha x_\alpha^k}
 \end{equation}
 where the $x_\alpha= \exp(-\beta_\alpha)$ counts the number of matrices of type $X_\alpha$ and $\beta_\alpha$ can be thought of as an effective temperature for the $X_\alpha$ degree of freedom.

 Again, the critical surface occurs when
 \begin{equation}
 \sum x_\alpha = 1
 \end{equation}
 and it is in this instance that $x_\alpha$ can also be interpreted as a single site probability for picking $X_\alpha$.
 
 For the microcanonical partition function, we fix $N_\alpha$, the number of $X_\alpha$. Call $E= \sum N_\alpha$. Similar arguments to the two matrix case show that 
 \begin{equation}
 Q_{multi} \simeq {E \choose N_1, \dots, N_s}\simeq \exp( E\log E -\sum N_\alpha \log N_\alpha)
 \end{equation}
 So that the entropy is 
 \begin{equation}
 S(N_\alpha) = E\log E -\sum N_\alpha \log N_\alpha
 \end{equation}
 The effective temperature for mode $\alpha$ is
 \begin{equation}
 \beta_\alpha = \log E - \log N_\alpha
 \end{equation} 
 It is then straightforward to check that
 \begin{equation}
 \exp(-\beta_\alpha) = N_\alpha/E = x_\alpha
 \end{equation}
 is the probability of finding an $X_\alpha$ on a single site, and that
 \begin{equation}
 \sum_\alpha \exp(-\beta_\alpha) = 1
 \end{equation} 
 matching the critical surface of the canonical ensemble.
 
 If all the $\beta_\alpha$ are equal to each other and equal to $\beta$, then we find that the critical value occurs at $\beta_{crit}= \log(s)$. Again, this matches the Hagedorn temperature for the system of $s$ matrices, all of them with
 the same mass.
 
 For any such model, the double trace deformation with negative coefficient produce the required negative specific heat. Similarly, generic non-planar interactions should render the specific heat negative in an intermediate energy regime.

 \section{Conclusion and outlook}\label{ref:con}
 
 In this paper I have argued that in the context of the gauge/gravity duality the negative specific heat associated to small black holes in the gravity dual of a gauge theory necessarily arises form non-planar 
 interactions. It is important that for this description one work in the microcanonical ensemble. 
 
 More importantly, the Hagedorn behavior of the free theory is already almost at a first order phase transition and can be perturbed to obtain negative specific heat without difficulty. 
 
 In this paper I studied a simple solvable double trace operator deformation of a two matrix model to show that if the double trace perturbation is attractive, the system has negative specific heat
 at intermediate temperatures. The negative specific heat can happen perturbatively below the energy at which the Hagedorn behavior ends. This end is non-perturbative in that it should be associated to a stringy exclusion principle. After this energy, in the free model, the system has  positive 
 specific heat.  In this same matrix model, I also showed that in the free theory the canonical and microcanonical ensemble description of the system agree with each other to describe the details of the
 Hagedorn behavior.  
 This analysis can be made to look like black hole thermodynamics of charged black holes. The negative specific effect can be characterized as being due to an effective redshift produced by gravitational attraction. There is also a bound on the charge given the energy, and the extremal configurations go 
 to zero temperature as one takes the limit, even when the double trace deformation is present.

  I also argued, based on what appear to be reasonable assumptions, that generic non-planar perturbations with characteristics of random matrices  lead to  negative specific heat
 without care about the sign of the interactions.  It is also in a way that is somewhat independent of the details of the interactions. 
 
 One can expect that this implies that most large $N$ dynamics, except for fine tuning will have configurations with negative specific heat. For field theories with a gap in more than $0+1$ dimensions, this should indicate that 
generically large $N$ theories have a first order phase transition. This should be interpreted as confinement/deconfinement. Configurations at fixed energy (and volume) at the transition temperature should therefore 
have phase separation in space. One can think of this as a thermodynamic instability of the uniform configuration. 

 The natural dual for this phase separation is to have localized black holes versus delocalized black holes. It is expected generically that the end-point of the Gregory-Laflamme instability \cite{Gregory:1993vy} at sufficiently large volumes will lead to localized black holes.
 The reason for the qualifier is that finite size effects can give rise to additional physics: one can expect that the deconfined phase at the critical temperature has some other correlation length scales associated to them, and if that length scale is larger than the size of the 
 deconfined droplet or the confined region the physics should be different. These would be the effects of the interface, which I have not studied in the models above, as they did not have an additional spatial dimension. It would be very interesting to explore these interfaces further.

 \acknowledgments
The author would like to thank D. Marolf and S. Ramgoolam for various discussions and L. Yaffe for some correspondence.
 Work  supported in part by the department of Energy under grant {DE-SC} 0011702.


\begin{thebibliography}{99}
 
\bibitem{Witten:1998zw} 
  E.~Witten,
``Anti-de Sitter space, thermal phase transition, and confinement in gauge theories,''
  Adv.\ Theor.\ Math.\ Phys.\  {\bf 2}, 505 (1998)
  doi:10.4310/ATMP.1998.v2.n3.a3
  [hep-th/9803131].
 
 
\bibitem{Hawking:1982dh} 
  S.~W.~Hawking and D.~N.~Page,
  ``Thermodynamics of Black Holes in anti-De Sitter Space,''
  Commun.\ Math.\ Phys.\  {\bf 87}, 577 (1983).
  doi:10.1007/BF01208266
 
 
\bibitem{Maldacena:2001kr} 
  J.~M.~Maldacena,
``Eternal black holes in anti-de Sitter,''
  JHEP {\bf 0304}, 021 (2003)
  doi:10.1088/1126-6708/2003/04/021
  [hep-th/0106112].
 
\bibitem{VanRaamsdonk:2010pw} 
  M.~Van Raamsdonk,
 ``Building up spacetime with quantum entanglement,''
  Gen.\ Rel.\ Grav.\  {\bf 42}, 2323 (2010)
  [Int.\ J.\ Mod.\ Phys.\ D {\bf 19}, 2429 (2010)]
  doi:10.1007/s10714-010-1034-0, 10.1142/S0218271810018529
  [arXiv:1005.3035 [hep-th]].
 
 
\bibitem{Berenstein:2018lrm} 
  D.~Berenstein,
  ``Submatrix deconfinement and small black holes in AdS,''
  JHEP {\bf 1809}, 054 (2018)
  doi:10.1007/JHEP09(2018)054
  [arXiv:1806.05729 [hep-th]].
 
\bibitem{Susskind:1993ws} 
  L.~Susskind,
  ``Some speculations about black hole entropy in string theory,''
  In *Teitelboim, C. (ed.): The black hole* 118-131
  [hep-th/9309145].


\bibitem{Horowitz:1996nw} 
  G.~T.~Horowitz and J.~Polchinski,
  ``A Correspondence principle for black holes and strings,''
  Phys.\ Rev.\ D {\bf 55}, 6189 (1997)
  doi:10.1103/PhysRevD.55.6189
  [hep-th/9612146].
 
 
\bibitem{Aharony:2003sx} 
  O.~Aharony, J.~Marsano, S.~Minwalla, K.~Papadodimas and M.~Van Raamsdonk,
  ``The Hagedorn - deconfinement phase transition in weakly coupled large N gauge theories,''
  Adv.\ Theor.\ Math.\ Phys.\  {\bf 8}, 603 (2004)
  doi:10.4310/ATMP.2004.v8.n4.a1
  [hep-th/0310285].
 
 
\bibitem{Sundborg:1999ue} 
  B.~Sundborg,
  ``The Hagedorn transition, deconfinement and N=4 SYM theory,''
  Nucl.\ Phys.\ B {\bf 573}, 349 (2000)
  doi:10.1016/S0550-3213(00)00044-4
  [hep-th/9908001].
 
\bibitem{Gross:1980he} 
  D.~J.~Gross and E.~Witten,
  ``Possible Third Order Phase Transition in the Large N Lattice Gauge Theory,''
  Phys.\ Rev.\ D {\bf 21}, 446 (1980).
  doi:10.1103/PhysRevD.21.446
 
\bibitem{Raha:2017jgv} 
  S.~Raha,
  ``Hagedorn temperature in superstring bit model and SU(N) characters,''
  Phys.\ Rev.\ D {\bf 96}, no. 8, 086006 (2017)
  doi:10.1103/PhysRevD.96.086006
  [arXiv:1706.09951 [hep-th]].





\bibitem{Curtright:2017pfq} 
  T.~L.~Curtright, S.~Raha and C.~B.~Thorn,
  ``Color Characters for White Hot String Bits,''
  Phys.\ Rev.\ D {\bf 96}, no. 8, 086021 (2017)
  doi:10.1103/PhysRevD.96.086021
  [arXiv:1708.03342 [hep-th]].



\bibitem{Beccaria:2017iqb} 
  M.~Beccaria,
  ``Thermal properties of a string bit model at large N,''
  JHEP {\bf 1710}, 200 (2017)
  doi:10.1007/JHEP10(2017)200
  [arXiv:1709.01801 [hep-th]].


\bibitem{Maldacena:1998bw} 
  J.~M.~Maldacena and A.~Strominger,
  ``AdS(3) black holes and a stringy exclusion principle,''
  JHEP {\bf 9812}, 005 (1998)
  doi:10.1088/1126-6708/1998/12/005
  [hep-th/9804085].

\bibitem{McGreevy:2000cw} 
  J.~McGreevy, L.~Susskind and N.~Toumbas,
  ``Invasion of the giant gravitons from Anti-de Sitter space,''
  JHEP {\bf 0006}, 008 (2000)
  doi:10.1088/1126-6708/2000/06/008
  [hep-th/0003075].

\bibitem{Balasubramanian:2001nh} 
  V.~Balasubramanian, M.~Berkooz, A.~Naqvi and M.~J.~Strassler,
  ``Giant gravitons in conformal field theory,''
  JHEP {\bf 0204}, 034 (2002)
  doi:10.1088/1126-6708/2002/04/034
  [hep-th/0107119].

\bibitem{Corley:2001zk} 
  S.~Corley, A.~Jevicki and S.~Ramgoolam,
``Exact correlators of giant gravitons from dual N=4 SYM theory,''
  Adv.\ Theor.\ Math.\ Phys.\  {\bf 5}, 809 (2002)
  doi:10.4310/ATMP.2001.v5.n4.a6
  [hep-th/0111222].


\bibitem{BR}
D.~Berenstein, S. Ramgoolam, {\em Work in progress}.


\bibitem{Caldarelli:2004mz} 
  M.~M.~Caldarelli, D.~Klemm and P.~J.~Silva,
  ``Chronology protection in anti-de Sitter,''
  Class.\ Quant.\ Grav.\  {\bf 22}, 3461 (2005)
  doi:10.1088/0264-9381/22/17/007
  [hep-th/0411203].

\bibitem{Milanesi:2005tp} 
  G.~Milanesi and M.~O'Loughlin,
  ``Singularities and closed time-like curves in type IIB 1/2 BPS geometries,''
  JHEP {\bf 0509}, 008 (2005)
  doi:10.1088/1126-6708/2005/09/008
  [hep-th/0507056].




\bibitem{Witten:2001ua} 
  E.~Witten,
  ``Multitrace operators, boundary conditions, and AdS / CFT correspondence,''
  hep-th/0112258.

\bibitem{Berkooz:2002ug} 
  M.~Berkooz, A.~Sever and A.~Shomer,
  ``'Double trace' deformations, boundary conditions and space-time singularities,''
  JHEP {\bf 0205}, 034 (2002)
  doi:10.1088/1126-6708/2002/05/034
  [hep-th/0112264].

\bibitem{Minahan:2002ve} 
  J.~A.~Minahan and K.~Zarembo,
  ``The Bethe ansatz for N=4 superYang-Mills,''
  JHEP {\bf 0303}, 013 (2003)
  doi:10.1088/1126-6708/2003/03/013
  [hep-th/0212208].

\bibitem{Harmark:2014mpa} 
  T.~Harmark and M.~Orselli,
  ``Spin Matrix Theory: A quantum mechanical model of the AdS/CFT correspondence,''
  JHEP {\bf 1411}, 134 (2014)
  doi:10.1007/JHEP11(2014)134
  [arXiv:1409.4417 [hep-th]].
\bibitem{Harmark:2016cjq} 
  T.~Harmark,
  ``Interacting Giant Gravitons from Spin Matrix Theory,''
  Phys.\ Rev.\ D {\bf 94}, no. 6, 066001 (2016)
  doi:10.1103/PhysRevD.94.066001
  [arXiv:1606.06296 [hep-th]].

\bibitem{Hanada:2016pwv} 
  M.~Hanada and J.~Maltz,
  ``A proposal of the gauge theory description of the small Schwarzschild black hole in AdS$_5\times$S$^5$,''
  JHEP {\bf 1702}, 012 (2017)
  doi:10.1007/JHEP02(2017)012
  [arXiv:1608.03276 [hep-th]].

\bibitem{Asplund:2008xd} 
  C.~T.~Asplund and D.~Berenstein,
  ``Small AdS black holes from SYM,''
  Phys.\ Lett.\ B {\bf 673}, 264 (2009)
  doi:10.1016/j.physletb.2009.02.043
  [arXiv:0809.0712 [hep-th]].


\bibitem{Gregory:1993vy} 
  R.~Gregory and R.~Laflamme,
  ``Black strings and p-branes are unstable,''
  Phys.\ Rev.\ Lett.\  {\bf 70}, 2837 (1993)
  doi:10.1103/PhysRevLett.70.2837
  [hep-th/9301052].

 \end{thebibliography}
\end{document}